\documentclass{article}

\usepackage{arxiv}

\usepackage[utf8]{inputenc}
\usepackage[T1]{fontenc}
\usepackage{hyperref}
\usepackage{url}
\usepackage{booktabs}
\usepackage{amsmath}
\usepackage{amsfonts}
\usepackage{nicefrac}
\usepackage{microtype}
\usepackage{cleveref}
\usepackage{graphicx}
\usepackage{natbib}
\usepackage{doi}
\usepackage{xspace}
\usepackage{listings}
\usepackage{xcolor}
\usepackage{subcaption}
\usepackage{multirow}
\usepackage{tikz}
\usetikzlibrary{arrows.meta,positioning,shapes.geometric,fit}

\newcommand{\code}[1]{\mbox{\texttt{#1}}}
\newcommand{\zerodep}{\mbox{zerodep}\xspace}

\lstdefinestyle{pythonstyle}{
    language=Python,
    basicstyle=\ttfamily\small,
    keywordstyle=\color{blue},
    commentstyle=\color{gray},
    stringstyle=\color{red!60!black},
    breaklines=true,
    frame=single,
    framesep=3pt,
    numbers=left,
    numberstyle=\tiny\color{gray},
    xleftmargin=2em,
}

\title{Stdlib or Third-Party?\\
Empirical Performance and Correctness\\
of LLM-Assisted Zero-Dependency Python Libraries}

\newif\ifuniqueAffiliation
\uniqueAffiliationfalse

\ifuniqueAffiliation
\author{
\href{https://orcid.org/0000-0001-8353-0821}{\includegraphics[scale=0.06]{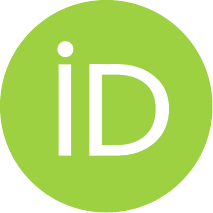}\hspace{1mm}Peng Ding} \\
University of Chicago \\
\texttt{dingpeng@uchicago.edu}
}
\else
\author{
\href{https://orcid.org/0000-0001-8353-0821}{\includegraphics[scale=0.06]{orcid.pdf}\hspace{1mm}Peng Ding} \\
University of Chicago \\
\texttt{dingpeng@uchicago.edu} \\
\And
Rick Stevens \\
University of Chicago \\
Argonne National Laboratory \\
\texttt{stevens@cs.uchicago.edu} \\
}
\fi

\renewcommand{\shorttitle}{Stdlib or Third-Party?}

\hypersetup{
pdftitle={Stdlib or Third-Party? Empirical Performance and Correctness of LLM-Assisted Zero-Dependency Python Libraries},
pdfsubject={Computer Science - Software Engineering},
pdfauthor={Peng Ding, Rick Stevens},
pdfkeywords={Python, standard library, zero dependency, LLM code generation, benchmarking, software engineering, empirical study},
}

\begin{document}
\maketitle

\begin{abstract}
Third-party Python libraries introduce dependency management overhead, supply chain
risk, and deployment friction in constrained environments. A natural question is
how much of this ecosystem can be replicated using only Python's standard library
--- and at what correctness and performance cost. We address this empirically
through \zerodep, a growing collection of single-file Python modules, each a
stdlib-only reimplementation of a popular third-party library, developed with
LLM assistance under strict constraints: no external imports, single file,
drop-in API compatibility, and mandatory correctness validation against the
reference library. Spanning over 40 modules across 12 categories --- including
serialization, networking, cryptography, agent protocols, and text processing ---
\zerodep provides a controlled testbed for two interrelated questions:
(1)~\emph{Where does the stdlib suffice?} and (2)~\emph{Can LLMs effectively
generate correct, performant code under tight symbolic constraints?}
Systematic benchmarking shows that stdlib-only implementations achieve
performance parity (within 2$\times$ of the reference) in the majority of
cases. The primary performance cliff is C-extension-backed computation
(image processing, binary serialization, low-level crypto), not the inherent
overhead of pure-Python third-party libraries. Conversely, many widely-used
libraries carry architectural overhead that LLM-generated stdlib reimplementations
avoid, yielding 5--115$\times$ speedups in several categories.
Beyond runtime performance, stdlib-only modules reduce installed footprint by
orders of magnitude (median 31\,KB vs.\ 700\,KB), import 2.2$\times$ faster in
cold-start scenarios, and eliminate 88 transitive dependencies maintained by 128
distinct PyPI accounts --- quantifiably reducing supply-chain attack surface.
Analysis of the development process across all 44 modules reveals that
specification completeness, not model capability, is a strong predictor of
generation success, and that contract-linked review reduces review volume by
$5\times$ while eliminating compatibility regressions.
\zerodep is open-source at \url{https://github.com/Oaklight/zerodep}.
\end{abstract}

\keywords{Python \and Standard Library \and Zero Dependency \and LLM Code Generation \and Empirical Study \and Benchmarking \and Software Engineering}

\section{Introduction}
\label{sec:intro}

Python's package ecosystem has grown to nearly 900,000 projects on PyPI, providing
ready-made solutions for nearly every software task. Yet this abundance comes at a
cost: each \code{pip install} introduces not just the declared package but a graph of
transitive dependencies. Recent empirical studies find that a single declared dependency
triggers a median of 8--10 additional transitive packages~\citep{arafat2025dependency},
and that over 70\% of code in the average Python project's dependency closure is
never actually executed~\citep{zhang2024bloated}. The consequences range from
prolonged installation times and version conflicts to supply-chain attacks and
vulnerability exposure in code paths that an application never exercises~\citep{drosos2024bloat}.

One common response to this overhead is the \emph{copy-paste pattern}:
for small, self-contained utilities (a YAML parser, a retry decorator, a dotenv
reader), developers copy the relevant source directly into their project
rather than adding a dependency. \citet{jahanshahi2025copy} show
that copy-based reuse is far more prevalent in open-source software than commonly
assumed, accounting for a significant fraction of code sharing in large ecosystems.
Ad-hoc copying, however, lacks discipline: there is no systematic way to discover
which libraries are suitable candidates, no correctness guarantee, and no performance
baseline.

LLM-assisted development adds a related problem. Code-generating LLMs frequently
hallucinate package names that do not exist, creating latent supply-chain risks when
generated code is deployed~\citep{spracklen2024hallucinations}. Constraining code
generation to Python's standard library (stdlib) eliminates this class of error
entirely: every import is guaranteed to exist and is pinned to the Python version,
not to an external registry.

These two observations motivate \zerodep: a collection of single-file
Python modules, each a stdlib-only reimplementation of a popular third-party library,
developed with LLM assistance under strict constraints. The collection is curated,
versioned, and benchmarked; any module can be dropped into a project with a single
CLI invocation and zero runtime dependencies. Correctness is maintained by using
the reference library as an automated oracle for systematic validation.

Our goal is not to argue that third-party dependencies are inherently undesirable.
Instead, we study a concrete software-engineering decision: when does eliminating a
third-party dependency preserve enough correctness and performance to be worth the
resulting gains in footprint, deployability, and trust surface? Framed this way,
\zerodep is an empirical testbed for characterizing the conditions under which
dependency elimination is a sound engineering choice.

This paper makes the following contributions:

\begin{enumerate}
  \item \textbf{A scoped empirical testbed for dependency elimination}: a curated
    collection of \textbf{44 stdlib-only Python modules} spanning 12 categories and
    three complexity tiers, each paired with a correctness test suite and a
    performance benchmark against its reference library (\S\ref{sec:testbed}).

  \item \textbf{A three-regime empirical characterization of replacement outcomes}: a benchmark study across all 44
    modules that identifies three performance regimes --- parity, architectural
    win, and C-extension cliff --- and explains the mechanism behind each
    (\S\ref{sec:eval}).

  \item \textbf{Boundary conditions for practical replacement}: systematic
    comparison of \zerodep module behavior against reference libraries across
    hundreds of test cases, showing where behavioral equivalence holds, where
    simplifications remain acceptable, and where native acceleration or broader
    API surface makes replacement a poor fit (\S\ref{sec:eval}).

  \item \textbf{Quantified deployment and trust-surface gains}: systematic
    measurement of import-time memory, installed size, cold-start latency, and
    eliminated dependency surface, showing that single-file stdlib modules reduce
    footprint by orders of magnitude, lower cold-start latency by a median of
    $2.2\times$, and remove 88 transitive packages maintained by 128 distinct
    PyPI accounts (\S\ref{sec:eval}).

  \item \textbf{Process evidence for constrained LLM-assisted development}: a
    quantitative study of LLM-assisted generation across complexity tiers,
    identifying recurring failure patterns, the role of specification completeness
    as a strong predictor of success, and the effect of contract-linked review on
    development outcomes (\S\ref{sec:dev-analysis}).
\end{enumerate}

We address three interrelated research questions:
\begin{description}
  \item[\textbf{RQ1}] \emph{Where does stdlib suffice?} Across correctness and
    performance dimensions, for which library categories and complexity tiers can
    a stdlib-only implementation replace a popular third-party library without
    meaningful loss?
  \item[\textbf{RQ2}] \emph{What are the deployment and supply-chain advantages?}
    How much do stdlib-only modules reduce resource footprint, cold-start latency,
    and supply-chain attack surface compared to the third-party libraries they
    replace?
  \item[\textbf{RQ3}] \emph{What determines success in LLM-assisted constrained
    generation?}
    Under strict symbolic constraints (stdlib-only, single-file, oracle-validated),
    what failure patterns recur, what role does specification completeness play,
    and where does human intervention remain necessary?
\end{description}

We scope \zerodep to \emph{utility and infrastructure} libraries: parsers,
serializers, HTTP clients, retry decorators, configuration loaders, and similar
building blocks. Compute-intensive frameworks (NumPy, PyTorch, audio processing)
are excluded because their value is inseparable from hardware-accelerated
native code.

Roughly two-thirds of the modules achieve performance within $2\times$ of the
reference library, and many outperform it by $5$--$115\times$. The performance
boundary is C-extension-backed computation, not the overhead of pure-Python
third-party libraries. Stdlib-only modules also reduce installed footprint
from a median 700\,KB to 31\,KB, import $2.2\times$ faster, and eliminate
88 transitive packages maintained by 128 distinct PyPI accounts.

For LLM-assisted generation, complexity tier predicts success: simple-tier
modules converge in one or two iterations, while subsystem-tier modules
require architectural guidance from the developer before LLM assistance
becomes productive.

\section{Related Work}
\label{sec:related}

\subsection{Dependency Bloat and Supply-Chain Risk}

Dependency management in package ecosystems has long been recognized as a distinct
and challenging problem~\citep{abate2012dependency}.
\citet{decan2017dependency} provide an early cross-ecosystem comparison of
dependency evolution in npm, CRAN, and RubyGems, establishing that transitive
dependency growth is a systemic phenomenon, while
\citet{kula2018developers} find that 81\% of studied projects keep dependencies
outdated, even after security advisories.
\citet{sotovalero2021bloated} conduct the first large-scale debloating analysis for
Maven, finding that 75\% of dependencies are bloated and proposing automated
removal tools.
\citet{drosos2024bloat} extend this line of work to Python, conducting a
fine-grained inter-project dependency analysis and finding that a substantial
fraction of imported package code is never executed, with bloated regions
disproportionately harboring security vulnerabilities.
\citet{zhang2024bloated} study ML system containers
and find systematic dependency bloat at both the OS package and Python package
levels. \citet{arafat2025dependency} quantifies the transitive amplification
effect across ten package ecosystems: a single declared dependency expands to a
median of 8--10 transitive packages. If a library's full feature set can be replaced
by a stdlib-only module, the entire transitive dependency subtree is eliminated.

From a security perspective, \citet{zimmermann2019smallworld} show that npm's
dense dependency graph creates a ``small world'' in which a single compromised
package can propagate to thousands of dependents.
\citet{ohm2020backstabber} catalog 174 real-world malicious packages across npm,
PyPI, and RubyGems, while \citet{duan2021supplychain} measure the practical
feasibility of typosquatting and dependency confusion attacks on package managers.
\citet{ladisa2023taxonomy} systematize the broader attack surface, cataloging
over 100 unique attack vectors across package registries and build pipelines.
\citet{spracklen2024hallucinations}
demonstrate that code-generating LLMs introduce a novel variant: \emph{package
hallucinations} --- references to non-existent packages --- which adversaries can
exploit by registering the hallucinated names on PyPI.
\citet{liu2025vulnerable}
study LLM framework repositories specifically, finding widespread vulnerable
transitive dependencies in the codebases that underpin LLM inference and training
pipelines. Stdlib-only modules sidestep both problems: they cannot hallucinate
imports, and they carry zero transitive dependency surface.

\subsection{Copy-Based Reuse and Vendoring}

Copy-based code reuse is a longstanding alternative to dependency-mediated
sharing~\citep{krueger1992reuse}.
\citet{jahanshahi2025copy} show that it is far
more prevalent than commonly assumed, accounting for a significant fraction of code
sharing patterns across large ecosystems --- making it a legitimate alternative to
package dependencies rather than an anti-pattern. The practice has industrial
precedent: Google's monorepo vendors all third-party code into the source
tree~\citep{potvin2016google}, and \citet{cox2019surviving} argues that vendoring
is a principled strategy for controlling supply-chain risk when dependencies are
treated as liabilities rather than free goods.
\zerodep sits in this space but adds structure: a curated, versioned, and tested
collection of modules with a CLI that resolves inter-module dependencies
automatically. Where the copy-based reuse studied by Jahanshahi et al.\ lacks
correctness guarantees, \zerodep accompanies each module with a reference-oracle
test suite.

\subsection{LLM-Assisted Library Reimplementation}

\citet{zhao2024commit0} introduce Commit0, a benchmark that tasks LLM
agents with reimplementing popular Python libraries from their API specifications and
an initial test suite, without any human editing of the generated output. Commit0
reports $\sim$80\% test pass rates for simple, small-surface-area libraries but
below 20\% for medium-complexity modules, with near-zero success on networking,
cryptography, and protocol libraries under fully automated, single-pass generation.
These failure modes, including missing edge cases, API mis-bindings, and subtle behavioral
mismatches, do not disappear with one-shot generation; \zerodep's tiered
complexity classification, iterative human--LLM co-design, and systematic
oracle-based validation are a direct response to that gap between benchmark
feasibility and production readiness.
\citet{zhang2026repozero} concurrently study fully automated LLM
generation of entire code repositories from specifications, extending the scope of
Commit0 to multi-file projects. \zerodep complements both by focusing on the
post-generation regime: measuring what iteratively refined, constrained LLM-assisted
implementations achieve in practice.

\citet{fan2024oracle} propose oracle-guided program selection from LLM
outputs at ISSTA 2024, using automated test suites to select among multiple LLM
candidates. \zerodep employs a similar philosophy: the reference library's behavior
is the ground-truth oracle~\citep{barr2015oracle}, but it applies that idea in an
iterative refinement loop rather than one-shot candidate selection. Recent work on
LLM self-repair~\citep{chen2023selfdebug, olausson2024selfrepair}, verbal
reinforcement~\citep{shinn2023reflexion}, and automated program repair with
LLMs~\citep{xia2023apr, jiang2023apr} has shown that feeding test failures
back into the model improves generation quality, though gains are bounded by the
model's ability to diagnose its own errors. \zerodep's workflow adopts the same
feedback structure but keeps the human in the loop for architectural decisions that
models handle poorly.

\subsection{LLM Code Generation and Library Knowledge}

The rapid growth of LLM-based code generation, from
Codex~\citep{chen2021codex} and MBPP~\citep{austin2021mbpp} through
Code Llama~\citep{roziere2023codellama} and
AlphaCode~\citep{li2022alphacode} to full
repository-level agents evaluated on SWE-bench~\citep{jimenez2024swebench},
has motivated studies of where models succeed and fail. Evaluation methodology
has itself evolved: \citet{liu2023evalplus} show that HumanEval's original test
suites are insufficient to catch many subtle bugs, while
BigCodeBench~\citep{zhuo2024bigcodebench} specifically targets code that calls
library APIs, the exact pattern \zerodep reimplements.
\citet{pearce2022copilot} find that roughly 40\% of Copilot-generated code
exhibits security vulnerabilities in high-risk scenarios, underscoring the
importance of automated validation.
\citet{wang2025deprecated} evaluate LLM code completion on deprecated
API usage, finding that models frequently generate code targeting API versions that
no longer exist. \citet{robbes2012deprecation} show that developers themselves
react slowly to API deprecation, compounding the staleness problem.
Python's stdlib provides a markedly more stable API surface than
third-party libraries: major behavioral changes are rare and well-documented across
versions. This stability is an underappreciated practical advantage of stdlib-only
implementations. \citet{yu2024codereval} benchmark LLMs on pragmatic code
generation tasks that require third-party library calls and find that models perform
better on tasks that require only stdlib imports, consistent with the
motivation for constraining generation to the standard library.

\subsection{Positioning}

\zerodep is not an LLM benchmark. Commit0 and RepoZero treat library reimplementation
as a way to measure model capability; \zerodep treats it as engineering. The
distinction matters: benchmark work measures generation success under zero-shot or
few-shot conditions; \zerodep measures correctness and performance of the
\emph{resulting} artifacts after iterative development. \zerodep also contributes
something neither benchmark provides: every module is benchmarked against its
reference library under identical conditions, producing a quantitative picture of
where stdlib-only implementations actually stand relative to the third-party
ecosystem.

\section{The \zerodep\ Testbed}
\label{sec:testbed}

\subsection{Design Constraints}
Every \zerodep\ module is implemented using only the Python standard library; no
third-party imports are permitted in module source files. Users can therefore copy
or vendor any module into their own projects without introducing new runtime
requirements. Each module resides in a single \code{.py} file and encapsulates its
entire functionality there, simplifying auditing and embedding workflows without
multi-file packaging complexity. Modules provide drop-in API compatibility with
established libraries (e.g., \code{yaml}, \code{retry}, \code{httpclient}),
enabling substitution in existing codebases. Correctness takes precedence over raw
performance: conformance tests against reference implementations are the primary
acceptance criterion, and benchmarks are measured only after functional equivalence
is established.

\subsection{Module Taxonomy}
\zerodep\ classifies its modules along two orthogonal dimensions: \emph{tier} and
\emph{category}. The tier dimension defines implementation scope and complexity.
\textbf{Simple} modules address a single, narrowly scoped concern (typically fewer than
200 lines of code), such as comment stripping, retry decoration, or dotenv
parsing. \textbf{Medium} modules span multiple related concerns (200--800 lines),
for example an HTML parser that handles tree construction, CSS selection, and
serialization. \textbf{Subsystem} modules exceed 800 lines and provide
protocol-level or system-oriented services, including full HTTP clients, agent communication
protocols, or binary serialization frameworks.

The category dimension divides modules into twelve functional areas:
\emph{network}, \emph{protocol}, \emph{serialization}, \emph{validation},
\emph{text}, \emph{config}, \emph{terminal}, \emph{crypto}, \emph{image},
\emph{process}, \emph{storage}, and \emph{devtools}. Serialization modules (YAML,
JSON, XML, protobuf) are benchmarked side-by-side; network modules (HTTP client,
HTTP server, WebSocket, SSE) form a separate performance cohort.

\subsection{Module Anatomy}
Each \zerodep\ module file begins with a PEP 723-style frontmatter block that
encodes metadata used for versioning, dependency resolution, and documentation
generation:

\begin{lstlisting}[style=pythonstyle, basicstyle=\ttfamily\scriptsize, caption={Example module frontmatter.}]
# /// zerodep
# version = "0.3.1"
# deps = ["httpclient"]
# tier = "subsystem"
# category = "serialization"
# ///
\end{lstlisting}

The \code{version} field follows Semantic Versioning (SemVer) and is managed
exclusively via the \code{zerodep bump} command, which computes a content hash
to detect actual changes and prevent spurious version increments. The \code{deps}
list declares inter-module dependencies by name; these correspond to other
single-file modules in the registry and are resolved automatically by the CLI.
The \code{tier} and \code{category} fields must match one of the predefined
taxonomy values. \code{manifest.json} is automatically generated by
\code{zerodep manifest} to record each module's metadata, content hash, and
dependency graph; it is never edited by hand.

\subsection{The \code{zerodep} CLI}

The \zerodep\ command-line interface handles module discovery, retrieval,
versioning, and dependency management. \code{zerodep~list} enumerates all available
modules by category and tier. \code{zerodep~info} prints frontmatter
fields, declared dependencies, and documentation notes without downloading anything.

To embed a module into a project, the user runs
\code{zerodep~add}, which copies the specified module files into the
current directory and recursively resolves any declared \code{deps}:
\begin{lstlisting}[basicstyle=\ttfamily\footnotesize,frame=none,numbers=none,xleftmargin=1em]
$ zerodep add yaml retry
\end{lstlisting}
This places \code{yaml.py} and \code{retry.py} into the working directory with no
further installation step required.
\code{zerodep~update} fetches the
latest version when upstream modules change. \code{zerodep~bump} determines which
modules have changed since their declared version and increments version fields
accordingly. \code{zerodep~dep\text{-}graph} visualizes inter-module relationships as a
directed acyclic graph, useful for auditing the transitive dependency footprint
before committing to a set of modules.

\section{Module Overview}
\label{sec:modules}

\zerodep currently comprises 44 modules across 12 categories and three complexity
tiers. Table~\ref{tab:modules} lists all modules with their tier, category,
reference library, and implementation size (lines of code, excluding tests).
Modules without a reference library entry are standalone implementations without
a direct third-party counterpart.

\begin{table}[htbp]
\centering
\caption{Module counts and representative references by category. Tier mix is
  reported as simple/medium/subsystem, corresponding to narrow utilities,
  multi-concern modules, and protocol- or system-level subsystems. The full
  module list is available in the anonymized replication package.}
\label{tab:modules}
\small
\begin{tabular}{lccp{0.34\columnwidth}}
\toprule
\textbf{Category} & \textbf{Count} & \textbf{Tier mix} & \textbf{Representative references} \\
\midrule
Network & 5 & 1/1/3 & httpx, websockets \\
Protocol & 6 & 1/3/2 & a2a-protocol, jsonrpcserver \\
Serialization & 7 & 2/3/2 & PyYAML, protobuf, commentjson \\
Validation & 3 & 1/2/0 & pydantic, packaging \\
Text \& Markup & 5 & 1/4/0 & beautifulsoup4, mistune \\
Configuration & 2 & 1/0/1 & python-dotenv, python-decouple \\
Terminal & 3 & 2/1/0 & tabulate \\
Crypto & 1 & 0/1/0 & PyCryptodome \\
Image & 2 & 1/1/0 & Pillow, qrcode \\
Process & 3 & 2/0/1 & tenacity, sh \\
Storage & 2 & 0/1/1 & cachetools, sqlitedict \\
DevTools & 5 & 1/2/2 & structlog, APScheduler \\
\bottomrule
\end{tabular}
\end{table}

\subsection{Category Overview}

The collection is deliberately skewed toward utility and infrastructure code rather
than compute-heavy frameworks. The largest subsystem modules appear in the network
and protocol categories, including \code{httpclient}, \code{httpserver}, \code{a2a},
and \code{acp}, each representing end-to-end interfaces rather than small helper
utilities. Serialization exhibits the widest performance range in the collection:
\code{yaml} and \code{jsonc} benefit from targeted implementations, whereas
\code{protobuf} illustrates the C-extension cliff. Text, validation, and
configuration modules occupy the middle ground, where stdlib implementations can
preserve core behavior but may omit the broader extension surface of their
reference libraries.
\code{aes} provides two implementation paths: a pure-Python AES core (for
environments without system OpenSSL) and an \code{openssl}-subprocess-backed path
that outperforms PyCryptodome by 1.5--8$\times$. \code{png} implements PNG
encoding and decoding using \code{zlib} and \code{struct}; performance is
7--45$\times$ below Pillow across payload sizes, as pixel-level processing
fundamentally benefits from SIMD-accelerated C code.

\paragraph{Process and Storage.}
\code{retry} reimplements \code{tenacity}'s decorator-based retry logic and is
36$\times$ faster in decorator overhead, as tenacity's general-purpose configuration
system adds substantial per-call cost. \code{scheduler} provides cron-expression
parsing and job scheduling via \code{threading}, outperforming \code{croniter} by
5--10$\times$ on next-fire-time computation. \code{cache} implements LRU, LFU,
and TTL caches matching the \code{cachetools} interface, achieving parity within
10--30\% across workloads.

\section{Methodology}
\label{sec:methodology}

\subsection{Correctness Evaluation}

Each \zerodep module is paired with a dedicated correctness test suite
(named \code{test\_correctness.py} per module) that compares module behavior
against the reference library on a shared set of inputs. This approach is an
instance of \emph{differential testing}~\citep{mckeeman1998differential}: the
reference library is the ground-truth oracle, and any divergence in output
constitutes a test failure. Test vectors cover
typical usage, edge cases (empty inputs, boundary values, malformed data), and
format-specific corner cases (e.g., YAML null representations, JSON comment
variants, protobuf field encoding rules). Tests are structured so that the same
input is fed to both the \zerodep module and the reference library, and their
outputs are compared for equality.

Import isolation is a recurring complication: several \zerodep modules share names
with their reference counterparts (e.g., \code{yaml.py} vs.\ PyYAML's \code{yaml}
package). Test files manage \code{sys.path} and \code{sys.modules} explicitly to
load both implementations in the same process without interference.

Correctness tests are run under \code{pytest} using the \code{conda} environment
\code{zerodep}, which includes all reference libraries pinned to specific versions
(listed in \code{pyproject.toml} under optional benchmark dependency groups).
All 44 modules pass their correctness suites; modules with intentional API
simplifications (e.g., \code{validate} omitting rarely-used Pydantic validators)
document the omissions explicitly in their module docstring.

\subsection{Performance Benchmarking}

Performance benchmarks use \code{pytest\text{-}benchmark}
with its default statistical model: each benchmark runs until the timing
estimate stabilizes (minimum 5 rounds, maximum determined by a 1-second
calibration target). Following best practices for rigorous performance
evaluation~\citep{georges2007benchmarking}, we report mean execution
time per operation and derived ops/second. All benchmarks run on the
same hardware (single machine, single-threaded unless the operation is
inherently concurrent), so that any timing difference reflects the
implementations rather than the environment.

Each benchmark test class contains two methods: one invoking the \zerodep
implementation (\code{test\_zerodep}) and one invoking the reference library
(\code{test\_reference}), with identical inputs. This structure ensures that
any variance in input construction or environment is shared equally between
the two implementations.

Benchmark results are collected into a \code{benchmark-results.json} file by
\code{pytest-benchmark} with JSON output and processed by a report generator
(\code{generate\_bench\_report.py}). All correctness tests and benchmarks
run automatically on every release via GitHub Actions, using the same
\code{pyproject.toml}-pinned reference library versions. Results are archived in an anonymized replication package prepared for review.
The full test suite, benchmark scripts, environment specification, and processed
benchmark outputs are available to the program committee in anonymized form,
enabling independent reproduction of all reported numbers while preserving the
double-anonymous review process.

\subsection{Performance Parity Definition}

We define \emph{performance parity} as a speed ratio $r = t_{\text{ref}} / t_{\zerodep}$
in the range $[0.5, 2.0]$, i.e., the \zerodep implementation is no more than
$2\times$ slower than the reference. This threshold reflects the nature of the
utility and infrastructure libraries in scope. For I/O-bound operations (HTTP
requests, file parsing, configuration loading), the library call itself is orders
of magnitude faster than the surrounding I/O; a $2\times$ difference at the
microsecond level is undetectable against millisecond-scale network or disk latency.
For CPU-bound utility operations such as serialization, modules in the parity range
either handle inputs small enough that absolute time differences are negligible, or
have already been identified as falling into the C-extension cliff (§\ref{sec:eval})
and classified as \emph{slower}. Modules with $r > 2.0$ are classified as
\emph{faster}; modules with $r < 0.5$ are classified as \emph{slower}. For modules
with multiple benchmark operations (e.g., different payload sizes or operation
types), we classify each operation independently and report the distribution.

\subsection{Resource Footprint Measurement}
\label{sec:method-footprint}

We measure three complementary dimensions of resource consumption that determine
suitability for deployment in constrained environments.

\paragraph{Import-time memory.}
Peak heap allocation during module import is measured via Python's
\code{tracemalloc}. Each measurement spawns a fresh subprocess that starts
\code{tracemalloc}, imports the module under test, and reports
\code{get\_traced\_memory()[1]} (peak bytes). For \zerodep modules, the subprocess
adds only the module's directory to \code{sys.path}; for reference libraries, the
subprocess runs inside a virtual environment with the library and its full
transitive dependency tree installed. Each module is measured 10 times; we
report median and interquartile range.  Because \code{tracemalloc} instruments
only CPython's memory allocator, it does not capture memory-mapped C-extension
shared objects; import-memory figures for C-extension-heavy references
(PyCryptodome, Pillow, protobuf) therefore \emph{underreport} their true footprint.

\paragraph{Installed footprint.}
For each reference library, we create a fresh virtual environment, install the
library at its pinned version (from \code{pyproject.toml}), and measure the total
\code{site-packages} size via \code{du~-sb}, subtracting the baseline size of an
empty virtual environment. The \zerodep counterpart is measured as the file size
of the single \code{.py} module. We also record the transitive dependency count
(\code{pip~list} entries minus \code{pip} and \code{setuptools}).

\paragraph{Cold-start latency.}
Import latency is measured in a fresh subprocess using
\code{time.perf\_counter\_ns}: the timer starts immediately before the
\code{import} statement and stops immediately after. Each module is measured
30 times; we report median, 5th, and 95th percentiles. This metric is relevant
for serverless and edge deployments where per-invocation cold-start cost
dominates.

All footprint measurements were collected on the same machine used for performance
benchmarks (Linux, Python~3.12), ensuring consistency across evaluation dimensions.

\subsection{Supply-Chain Surface Measurement}
\label{sec:method-supplychain}

To quantify the supply-chain attack surface that \zerodep eliminates, we measure
four properties of each reference library's transitive dependency tree:

\begin{enumerate}
  \item \textbf{Transitive dependency count}: the number of packages installed by
    \code{pip~install} in a fresh virtual environment, excluding \code{pip} and
    \code{setuptools}.
  \item \textbf{Third-party lines of code}: the total line count of all
    \code{.py} files across every installed package, enumerated via
    \code{pip~show~-{}-files}.
  \item \textbf{Known vulnerabilities}: \code{pip-audit} run against the installed
    environment, querying the Open Source Vulnerability (OSV) database.
  \item \textbf{Maintainer count}: the number of distinct PyPI maintainer accounts
    across all transitive dependencies, collected from the PyPI JSON~API. Each
    maintainer account represents a trust point whose compromise could inject
    code into downstream consumers.
\end{enumerate}

For \zerodep modules, all four metrics are zero by construction: stdlib-only
modules carry no third-party imports, no third-party code, no third-party
vulnerability surface, and no third-party maintainers.

\subsection{LLM-Assisted Development Workflow}
\label{sec:dev-workflow}

Each \zerodep module is developed through an iterative human--LLM collaboration
process organized around five activities.

\paragraph{Specification.}
The developer writes a behavioral contract that defines the public API surface,
required behavior, error semantics, intentional exclusions, and the role (if any)
of a reference library.
For protocol modules (e.g., \code{a2a}, \code{acp}), the protocol specification
is the primary input. The contract is frozen before generation begins;
changes discovered during implementation are recorded as specification concerns
and resolved by the developer rather than silently adopted.

\paragraph{Separated generation and testing.}
An LLM (primarily Claude and o-series models) generates an initial implementation
subject to the \zerodep constraints: stdlib-only, single file, matching the
target API. A separate session then generates correctness tests from the
specification alone, without access to the implementation. This separation
prevents tests from over-fitting to implementation details or encoding the same
misunderstandings as the code --- a concern documented in LLM test
generation~\citep{schafer2024coverup}.

\paragraph{Iterative validation and repair.}
The implementation is validated against both the generated tests and the
human-authored reference-library oracle~\citep{barr2015oracle}. Test failures and static-analysis
violations are fed back to the LLM for bounded repair iterations. Per-test
timeouts and attempt budgets prevent unbounded loops; exhaustion halts the
attempt and retains partial artifacts for diagnosis.

\paragraph{Contract-linked review and correction.}
Review findings are tied to specification clauses rather than left as general
suggestions. This prevents \emph{review drift}, in which reviewers suggest
behavioral changes beyond the original scope. Address rounds are bounded;
corrections that pass some checks while breaking others trigger rollback to
the last passing checkpoint rather than compounding regressions.

\paragraph{Human review and release.}
The developer reviews the final implementation, resolves remaining edge cases,
and bumps the module version. Scope decisions, ambiguous specification points,
and disputed behavioral findings remain human responsibilities throughout. LLMs
are used only to propose candidate implementations, tests, and bounded repairs;
study design, experiment configuration, result interpretation, and final
acceptance of code and artifacts remain author responsibilities. All reported
quantitative results come from scripted executions and human-reviewed artifacts.

Section~\ref{sec:dev-analysis} reports quantitative development outcomes and
recurring failure patterns across all 44 modules.

\section{Evaluation}
\label{sec:eval}

We evaluate all 44 \zerodep modules along four dimensions: behavioral correctness
and runtime performance (RQ1), resource footprint and deployment characteristics
(RQ2), supply-chain attack surface (RQ2), and LLM-assisted development
observations (RQ3, §\ref{sec:discussion}). All benchmarks were collected on a
single machine running Linux with Python 3.12 and the pinned reference library
versions listed in \code{pyproject.toml}.

\subsection{RQ1: Correctness}

All 44 modules pass their correctness suites, which cover typical usage, boundary
conditions, and format-specific edge cases. The main challenge is subtle behavioral
divergence rather than API absence; the oracle-based tests force \zerodep to match
reference behavior on cases such as YAML nulls, JSON comment handling, and protobuf
field ordering.

\subsection{RQ1: Performance Parity Overview}

Table~\ref{tab:perf-summary} summarizes the performance verdict for each module
across its primary benchmark operations. We classify modules into three regimes:
\emph{Faster} ($r > 2.0$, i.e., \zerodep is more than $2\times$ faster than the
reference), \emph{Parity} ($r \in [0.5, 2.0]$), and \emph{Slower} ($r < 0.5$).

\begin{table}[htbp]
\centering
\caption{Performance summary by regime.}
\label{tab:perf-summary}
\scriptsize
\setlength{\tabcolsep}{2pt}
\begin{tabular}{p{0.17\columnwidth}p{0.22\columnwidth}p{0.15\columnwidth}c}
\toprule
\textbf{Regime} & \textbf{Examples} & \textbf{Ratio} & \textbf{N} \\
\midrule
Arch. win & \code{jsonc}, \code{yaml} & $2$--$115\times$ & 19 \\
Parity & \code{dotenv}, \code{xml} & $0.5$--$2\times$ & 11 \\
C-ext. cliff & \code{protobuf}, \code{png} & $0.003$--$0.27\times$ & 6 \\
Mixed & \code{a2a}, \code{acp} & workload-dependent & 8 \\
\bottomrule
\end{tabular}
\setlength{\tabcolsep}{6pt}
\end{table}

Across the collection, 19 modules are consistently faster than their reference,
11 achieve parity, 6 are consistently slower, and the remainder depend on the
operation being measured. The main divide is not stdlib versus third-party per se,
but architectural overhead versus native acceleration.

\subsection{RQ1: The C-Extension Performance Cliff}

The consistently slower modules share a common characteristic: their reference
libraries rely on compiled C or Rust extensions for their performance-critical
paths.

\paragraph{Cryptography.}
\code{aes} in pure-Python mode is 300--17,000$\times$ slower than PyCryptodome
across cipher modes and payload sizes, a consequence of AES's byte-level
operations mapping poorly to Python's interpreter overhead. The \zerodep
\code{aes} module mitigates this through an alternative code path that delegates
to the system's OpenSSL installation via a subprocess call, achieving
1.5--8$\times$ \emph{faster} performance than PyCryptodome by eliminating the
Python object overhead on the hot path. The same pattern,
delegating to system libraries via stdlib subprocess or \code{ctypes},
applies to other computationally intensive tasks.

\paragraph{Binary serialization.}
\code{protobuf} is 6--72$\times$ slower than
\code{google\text{-}protobuf}'s C~extension. Protobuf encoding is inherently
byte-level work (varints, wire-type dispatch, field scanning), and the
pure-Python implementation using \code{struct} adds substantial per-field
overhead. Unlike AES, there is no system-library shortcut for arbitrary
protobuf schemas.

\paragraph{Image processing.}
\code{png} is 7--45$\times$ slower than Pillow for both encoding and decoding.
PNG processing involves row-level filter computation and zlib compression across
potentially millions of pixels, work that SIMD-accelerated C code handles orders
of magnitude faster than Python loops. The stdlib \code{zlib} module handles
compression but the pixel transformation loops remain in Python.

\paragraph{Confirming interpreter overhead via JIT and profiling.}
To verify that the performance cliff reflects CPython interpreter overhead rather
than algorithmic inefficiency, we ran the \zerodep implementations of the six
slower modules under PyPy~3.10 (which applies JIT compilation to Python bytecode)
alongside four faster/parity modules as controls. For the treatment group,
PyPy's JIT closes a substantial fraction of the gap: \code{protobuf} serialization
accelerates $12$--$20\times$, and \code{validate} reaches $4$--$5\times$ speedup,
while control-group modules (already fast under CPython) show only modest
improvement. This asymmetry confirms that the performance cliff is dominated by
per-operation interpreter overhead in tight inner loops, not by suboptimal
algorithms.

\code{cProfile} analysis of the three worst-case modules corroborates this.
In \code{aes} (pure-Python), 95\% of execution time concentrates in eight
byte-level functions (\code{\_mix\_single\_column}, \code{\_add\_round\_key},
\code{\_sub\_bytes}, \code{\_xtime}). In \code{png}, 94\% concentrates in six
pixel-level functions (\code{\_filter\_row}, \code{\_paeth}, filter scoring).
In \code{protobuf}, 83\% concentrates in four varint/encoding functions
(\code{\_write\_varint}, field encoders, \code{bytearray.append}).
In every case the hot path performs per-byte or per-element arithmetic that
maps directly to a handful of machine instructions in C but requires full
interpreter dispatch in CPython.

Table~\ref{tab:profiling} summarizes the hot-path concentration for each
profiled module.

\begin{table}[htbp]
\centering
\caption{Hot-path concentration in the three worst-case modules.}
\label{tab:profiling}
\scriptsize
\setlength{\tabcolsep}{3pt}
\begin{tabular}{lp{0.28\columnwidth}rr}
\toprule
\textbf{Module} & \textbf{Top function} & \textbf{Time} & \textbf{\%} \\
\midrule
\multirow{4}{*}{aes}
  & \code{\_mix\_single\_column} & 5.27\,s & 28.1 \\
  & \code{\_add\_round\_key}     & 3.31\,s & 17.6 \\
  & \code{\_sub\_bytes}          & 2.50\,s & 13.3 \\
  & \code{\_xtime}               & 2.48\,s & 13.2 \\
  & \multicolumn{2}{r}{\emph{Top 8 total:}} & \emph{95.1} \\
\midrule
\multirow{4}{*}{png}
  & \code{\_filter\_row}         & 1.42\,s & 32.4 \\
  & \code{\_paeth}               & 0.92\,s & 20.9 \\
  & filter \code{<genexpr>}      & 0.70\,s & 16.0 \\
  & \code{sum}                   & 0.51\,s & 11.6 \\
  & \multicolumn{2}{r}{\emph{Top 6 total:}} & \emph{93.8} \\
\midrule
\multirow{4}{*}{protobuf}
  & \code{\_write\_varint}       & 0.62\,s & 40.2 \\
  & int32 \code{<lambda>}        & 0.33\,s & 21.7 \\
  & \code{\_encode\_packed}      & 0.17\,s & 11.4 \\
  & \code{bytearray.append}      & 0.15\,s & 10.1 \\
  & \multicolumn{2}{r}{\emph{Top 4 total:}} & \emph{83.3} \\
\bottomrule
\end{tabular}
\setlength{\tabcolsep}{6pt}
\end{table}

\subsection{RQ1: Architectural Overhead as Opportunity}

Many widely-used third-party libraries are \emph{much slower} than targeted
stdlib implementations, despite having no C-extension dependency of their own.

When a reference library carries architectural overhead unrelated to the core
task, a targeted stdlib implementation can be much faster. The clearest
examples are \code{jsonc} ($75$--$115\times$), \code{httpclient}
($18$--$32\times$), \code{yaml} ($6$--$7\times$), and \code{soup}
($2.1$--$3.3\times$). The gains come from avoiding plugin layers,
intermediate representations, and transport abstractions that the narrower
implementation does not need.

\subsection{RQ2: Resource Footprint and Deployment Characteristics}

Beyond correctness and runtime speed, constrained deployment also rewards small
installed size, low import cost, and reduced transitive dependency surface.

\paragraph{Installed footprint.}
Table~\ref{tab:footprint} compares the on-disk footprint of each \zerodep module
(a single \code{.py} file) against the installed size of its reference library
including all transitive dependencies.

\begin{table*}[t]
\centering
\caption{Installed footprint: \zerodep module size vs.\ reference library tree.}
\label{tab:footprint}
\footnotesize
\setlength{\tabcolsep}{3pt}
\begin{tabular}{lrp{0.13\textwidth}rrl|@{\hspace{0.5em}}lrp{0.13\textwidth}rrl}
\toprule
\textbf{Mod.} & \textbf{ZD} & \textbf{Ref.} & \textbf{Size} & \textbf{D} & \textbf{R}
& \textbf{Mod.} & \textbf{ZD} & \textbf{Ref.} & \textbf{Size} & \textbf{D} & \textbf{R} \\
 & \textbf{\scriptsize KB} & & & & &  & \textbf{\scriptsize KB} & & & & \\
\midrule
s3           & 31  & boto3          & 24.6\,MB & 7 & 815$\times$
& soup         & 34  & bs4 & 1.4\,MB  & 3 & 41$\times$  \\
readability  & 36  & readability& 15.7\,MB & 5 & 443$\times$
& httpclient   & 96  & httpx          & 3.7\,MB  & 7 & 39$\times$  \\
aes          & 36  & pycryptodome   & 8.6\,MB  & 1 & 248$\times$
& scheduler    & 37  & croniter       & 1.3\,MB  & 3 & 35$\times$  \\
jsonrpc      & 14  & jsonrpcsrv  & 3.2\,MB  & 8 & 230$\times$
& protobuf     & 76  & protobuf       & 2.2\,MB  & 1 & 30$\times$  \\
validate     & 41  & pydantic       & 9.1\,MB  & 5 & 227$\times$
& structlog    & 28  & structlog      & 482\,KB  & 1 & 17$\times$  \\
frontmatter  & 14  & frontmatter & 3.1\,MB  & 2 & 219$\times$
& markdown     & 28  & mistune        & 454\,KB  & 1 & 16$\times$  \\
sse          & 21  & httpx          & 3.7\,MB  & 7 & 183$\times$
& retry        & 17  & tenacity       & 184\,KB  & 1 & 11$\times$  \\
acp          & 65  & agent-client& 9.6\,MB  & 6 & 152$\times$
& tabulate     & 28  & tabulate       & 248\,KB  & 1 & 9$\times$   \\
a2a          & 71  & a2a-proto   & 9.2\,MB  & 7 & 133$\times$
& dotenv       & 15  & dotenv  & 110\,KB  & 1 & 7$\times$   \\
yaml         & 36  & PyYAML         & 3.0\,MB  & 1 & 86$\times$
& runner       & 46  & sh             & 256\,KB  & 1 & 6$\times$   \\
httpserver   & 55  & flask          & 4.5\,MB  & 7 & 84$\times$
& cache        & 30  & cachetools     & 136\,KB  & 1 & 5$\times$   \\
ratelimit    & 37  & limits         & 2.9\,MB  & 5 & 81$\times$
& diff         & 22  & unidiff        & 66\,KB   & 1 & 3$\times$   \\
jsonschema   & 42  & jsonschema     & 2.9\,MB  & 6 & 70$\times$
& xml          & 27  & xmltodict      & 66\,KB   & 1 & 2$\times$   \\
semver       & 12  & packaging      & 721\,KB  & 1 & 59$\times$
& config       & 24  & decouple& 39\,KB   & 1 & 2$\times$   \\
\midrule
\multicolumn{12}{l}{\textbf{Aggregate:} median ZD 31\,KB; median ref 700\,KB; deps eliminated 128} \\
\bottomrule
\end{tabular}
\setlength{\tabcolsep}{6pt}
\end{table*}

\begin{figure}[t]
\centering
\includegraphics[width=0.92\columnwidth]{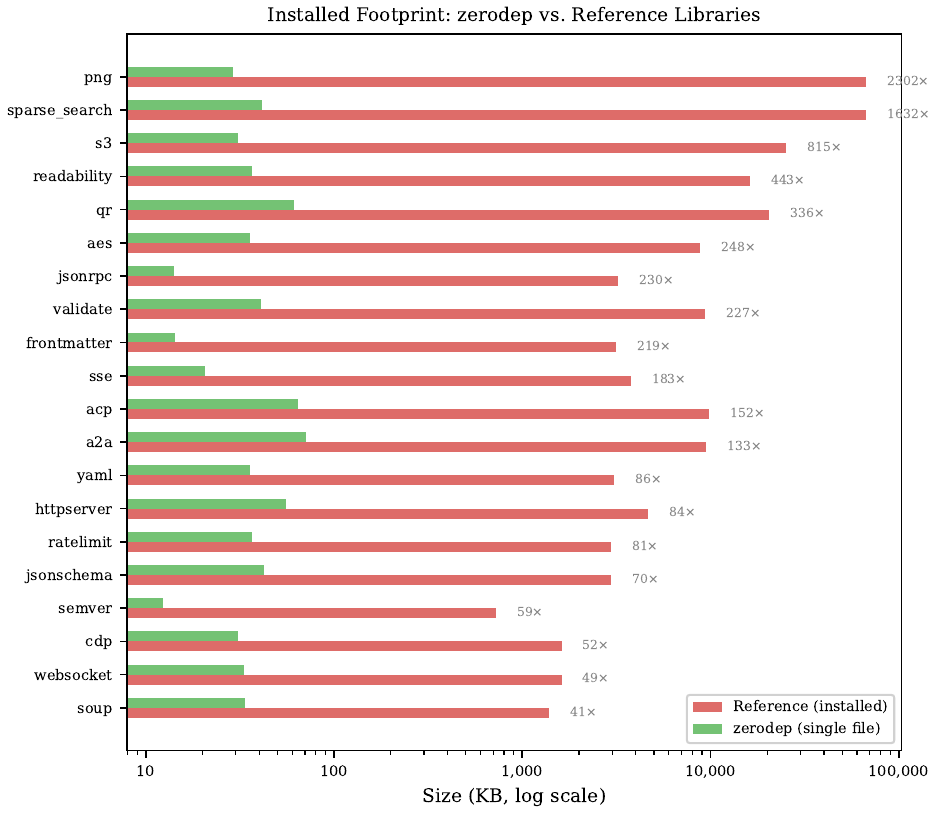}
\caption{Installed footprint comparison (log scale). Green bars show \zerodep
  single-file size; red bars show reference library installed size including
  transitive dependencies. Annotations show the reduction ratio.}
\label{fig:footprint}
\end{figure}

Figure~\ref{fig:footprint} visualizes the footprint gap on a logarithmic scale.

The median \zerodep module occupies 31\,KB; the median reference library installs
700\,KB including its transitive dependencies. Across all 39 modules with
reference counterparts, \zerodep eliminates 128 transitive dependencies. The
most dramatic reductions occur for libraries with heavy native components
(boto3: 815$\times$) or deep dependency trees (jsonrpcsrv: 230$\times$,
8 transitive deps).

\paragraph{Cold-start latency.}
Table~\ref{tab:coldstart} reports import latency in a fresh Python process,
a significant component of serverless cold-start cost.

\begin{table}[htbp]
\centering
\caption{Cold-start import latency summary.}
\label{tab:coldstart}
\scriptsize
\setlength{\tabcolsep}{3pt}
\begin{tabular}{p{0.22\columnwidth}p{0.16\columnwidth}c}
\toprule
\textbf{Group} & \textbf{Speedup} & \textbf{N} \\
\midrule
Largest gains & $4.4$--$10.4\times$ & 6 \\
Moderate gains & $2.0$--$3.4\times$ & 12 \\
Near parity & $1.0$--$1.7\times$ & 14 \\
\bottomrule
\end{tabular}
\setlength{\tabcolsep}{6pt}
\end{table}

\zerodep modules import a median of $2.2\times$ faster than their reference libraries. Import-time memory shows the same pattern, with the median \zerodep module allocating about $1.5\times$ less peak heap memory during import. Runtime memory measurements on six representative modules follow the same direction, indicating that much of the savings comes from architectural simplification rather than different algorithms.

\subsection{RQ2: Supply-Chain Attack Surface}

Every third-party dependency introduces code maintained by external parties into
the deployed application. Table~\ref{tab:supplychain} quantifies the aggregate
attack surface of the reference libraries that \zerodep replaces.

\begin{table}[htbp]
\centering
\caption{Supply-chain attack surface: aggregate metrics across all 39 reference
  libraries vs.\ \zerodep.}
\label{tab:supplychain}
\small
\begin{tabular}{lrr}
\toprule
\textbf{Metric} & \textbf{Reference libraries} & \textbf{\zerodep} \\
\midrule
Unique transitive dependencies    & 88        & 0 \\
Third-party lines of code         & 1,458,711 & 0 \\
Distinct PyPI maintainer accounts & 128       & 0 \\
Known CVEs (at pinned versions)   & 0         & 0 \\
\bottomrule
\end{tabular}
\end{table}

\begin{figure}[t]
\centering
\includegraphics[width=\columnwidth]{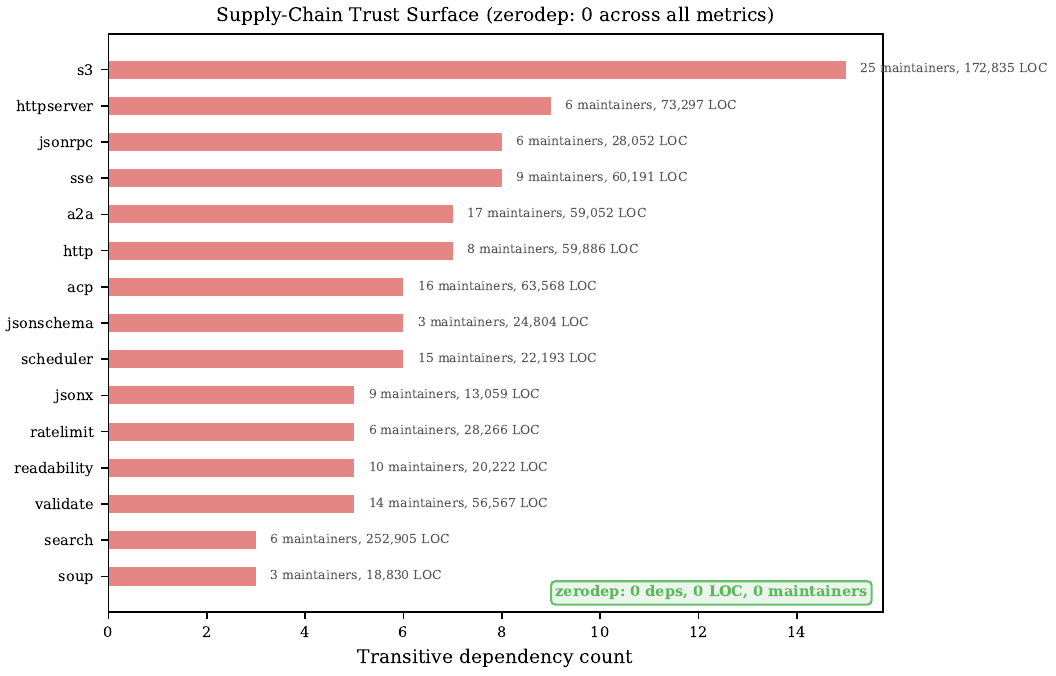}
\caption{Supply-chain trust surface per reference library: transitive dependency
  count (bars), with maintainer count and third-party LOC annotated.
  zerodep is zero across all metrics.}
\label{fig:supplychain}
\end{figure}

Figure~\ref{fig:supplychain} visualizes the per-module trust surface.
The 39 reference libraries collectively introduce 88 unique transitive packages
maintained by 128 distinct PyPI accounts, comprising 1.46 million lines of
third-party code.

Each maintainer account is a trust point. Recent incidents such as the
xz-utils backdoor and the polyfill.io takeover show that maintainer compromise is
not theoretical. The deepest trust surfaces in our set are \code{boto3}+\code{minio}
(15 deps, 25 maintainers), the Flask-family server stack (9 deps),
\code{httpx\text{-}sse}+\code{httpx} and \code{jsonrpcsrv} (8 deps), and
\code{pydantic} (5 deps, 14 maintainers). Zero pinned-version CVEs is incidental;
our main point is structural: \zerodep collapses the trust boundary to the Python
standard library alone.

\section{RQ3: LLM-Assisted Development Analysis}
\label{sec:dev-analysis}

The workflow described in §\ref{sec:dev-workflow} was applied to all 44 modules. This section summarizes the main process outcomes and failure patterns.

\subsection{Development Effort by Complexity Tier}

Table~\ref{tab:dev-effort} summarizes development effort by complexity tier,
reconstructed from repository issues, pull requests, commits, CI results, and
review records.

\begin{table}[htbp]
\centering
\caption{Development effort by complexity tier.}
\label{tab:dev-effort}
\small
\begin{tabular}{lrrr}
\toprule
\textbf{Metric} & \textbf{Simple} & \textbf{Medium} & \textbf{Subsystem} \\
\midrule
Modules                    & 14       & 17       & 13       \\
Median LLM iterations      & 1--3     & 2--5     & 5--10+   \\
Median validation attempts & 1--2     & 2--3     & 3--5     \\
Human architecture needed  & Rarely   & Sometimes & Always  \\
Typical wallclock (min)    & $\sim$10 & $\sim$20 & $>$60    \\
\bottomrule
\end{tabular}
\end{table}

Simple-tier modules typically converge in one to three LLM iterations, medium-tier modules require more repair for behavioral mismatches, and subsystem-tier modules still require substantial human architectural input before LLM assistance becomes productive. In practice, the LLM accelerates implementation rather than driving design for subsystem-tier work.

Review tends to increase LOC through error handling and edge-case guards, while refactoring partially reverses that growth; final modules typically remain within about $\pm 15\%$ of their initial LOC.

\subsection{Recurring Failure Patterns}

Across the 44 modules, six failure patterns recurred: specification ambiguity,
unverified corrections, review-induced regression, evaluation mismatch,
unbounded execution, and malformed structured outputs. The most important fix
was to freeze behavioral contracts before generation and route specification
questions to human adjudication. Combined with automated validation, bounded
repair attempts, and contract-linked review, this removed the most persistent
sources of regression and false completion.

\subsection{Specification Completeness as Success Predictor}

We evaluated LLM-generated implementations against held-out human-authored
test suites that were unavailable during generation. A clear gradient emerged:
modules with tighter specifications passed more held-out tests.

\begin{table}[htbp]
\centering
\caption{Compatibility with held-out human test suites, grouped by
  specification completeness. Held-out suites were unavailable during
  generation.}
\label{tab:spec-gradient}
\small
\renewcommand{\arraystretch}{1.15}
\begin{tabular}{p{0.22\columnwidth}p{0.14\columnwidth}p{0.56\columnwidth}}
\toprule
\textbf{Pattern} & \textbf{Modules} & \textbf{Interpretation} \\
\midrule
Complete agreement &
  \code{yaml} 67/67; \code{dotenv} 53/56 &
  Standardized or tightly specified APIs; remaining gaps are behavioral edge
  cases. \\
\addlinespace
Behavioral divergence &
  \code{markdown} 80/110 &
  Shared core API with differences on ambiguous or extended syntax not
  constrained by specification. \\
\addlinespace
API design divergence &
  \code{retry} 7/36; \code{filelock} 0/13 &
  Specifications permitted different parameter names, types, or interaction
  patterns; each implementation was valid against its own contract. \\
\addlinespace
Import-level mismatch &
  33 of 40 modules &
  Human test suites required export names not constrained by the generated
  specifications. \\
\bottomrule
\end{tabular}
\end{table}

The 33-of-40 import-level mismatch is a metadata omission, not a behavioral
defect: most specifications did not constrain public export names.
Modules with constrained APIs (\code{yaml}, \code{dotenv}) achieved
near-perfect held-out compatibility. Modules with open naming
(\code{retry}, \code{filelock}) diverged on surface design even though
each implementation was valid against its own contract.

For this class of task, specification engineering mattered more than model
selection. Adding API constraints, error semantics, and exclusions to a
contract improved held-out compatibility more than switching to a stronger
model.

\subsection{Review Drift and Its Resolution}

Unconstrained review was a persistent source of regressions during development.
When reviewers were given the implementation diff without a frozen specification,
they suggested behavioral changes beyond the original scope. Address rounds that
applied these suggestions sometimes improved internal benchmarks while reducing
compatibility with the reference library's behavior.

Table~\ref{tab:review-drift} compares review outcomes before and after adopting
specification-linked review.

\begin{table}[htbp]
\centering
\caption{Review behavior before and after contract-linked review.
  Metrics from calibration runs on matched modules.}
\label{tab:review-drift}
\small
\begin{tabular}{lrr}
\toprule
\textbf{Metric} & \textbf{Before} & \textbf{After} \\
\midrule
Review findings per module          & 19--34 & 4--5 \\
Blockers per module                 & 7      & 0--1 \\
External compatibility change       & $-5.5$\,pp & 0\,pp \\
Specification concerns raised       & n/a    & 0 \\
\bottomrule
\end{tabular}
\end{table}

Contract-linked review reduced review volume by approximately $5\times$ and
eliminated compatibility regressions (Table~\ref{tab:review-drift}). Reviewers
could flag issues tied to specification clauses or code quality, but not suggest
behavioral changes. Potential specification deficiencies were escalated to the
developer rather than applied by the model.

LLMs will accept any suggestion that lets the tests pass, even changes that
move behavior away from the intended design. A frozen specification prevents
this by defining ``correct'' independent of reviewer opinion or test-suite
coverage.

\section{Discussion}
\label{sec:discussion}

\subsection{The C-Extension Performance Cliff}

Our benchmarks reveal a pronounced performance cliff for pure-Python
implementations of interfaces typically backed by C extensions.
Among the 44 modules in \zerodep,
roughly 30 achieve within-$2\times$ parity compared to their reference
libraries, but C-extension workloads diverge sharply.
The PNG encoder is 7--45$\times$ slower than Pillow, the
Protobuf serializer runs 6--72$\times$ slower than
\code{google\text{-}protobuf}, and the pure-Python AES cipher
trails PyCryptodome by
300--17{,}000$\times$. These gaps reflect a well-known property of
interpreted Python: cross-language benchmarks show CPython trailing
compiled languages by one to two orders of magnitude on CPU-bound
workloads~\citep{pereira2017energy}, and the gap compounds in tight
loops over byte-level operations.

Our PyPy JIT experiments (\S\ref{sec:eval}) confirm that this cliff is dominated
by interpreter overhead rather than algorithmic weakness: JIT compilation
closes a substantial fraction of the gap for treatment-group modules while
control-group modules see minimal change. Profiling corroborates this: in every
case, over 80\% of execution time concentrates in fewer than eight functions
performing per-byte or per-element arithmetic.

Delegating cryptographic primitives to an external C library via \code{openssl}
subprocess invocations not only closes the gap but surpasses the reference: AES
encryption is 1.5--8$\times$ faster than PyCryptodome. Offloading hot loops to
native executables mitigates the performance cliff without adding non-stdlib
dependencies. The trade-off includes process-spawn overhead and platform-specific
filesystem interactions, which must be managed carefully in latency-sensitive contexts.

\subsection{Architectural Overhead as Hidden Cost}

Architectural bloat in pure-Python reference libraries accounts for more of the
collection's performance variation than C extensions do. Across the 19 modules
classified as ``faster'' (Table~\ref{tab:perf-summary}), the speedups come not
from a better algorithm but from removing indirection: plugin dispatch, intermediate
ASTs, transport abstractions, and configuration layers that the narrower stdlib
implementation does not need. Where the reference library is already I/O-bound
or string-only (\code{dotenv}, frontmatter, QR), parity is the natural outcome
and stdlib replacement has no performance motivation, only a packaging one.

\subsection{Deployment and Supply-Chain Advantages}

\zerodep also reduces deployment friction: the median module is a single 31\,KB
file, whereas the median reference library installs 700\,KB plus transitive
dependencies. The median import speedup is $2.2\times$, and replacing the 39
reference libraries removes 88 transitive packages and 128 maintainer trust
points. Dependency elimination is not a complete security solution, but it is a
structural reduction of packaging and supply-chain exposure.

\subsection{Practical Guidance}

Table~\ref{tab:decision-guide} summarizes the main decision patterns that emerge
from our study.

\begin{table}[htbp]
\centering
\caption{Decision guide for stdlib-only replacement.}
\label{tab:decision-guide}
\footnotesize
\setlength{\tabcolsep}{2pt}
\begin{tabular}{p{0.18\columnwidth}p{0.23\columnwidth}p{0.23\columnwidth}}
\toprule
\textbf{Case} & \textbf{Use stdlib-only} & \textbf{Keep dependency} \\
\midrule
Workload & Utility logic & Native hot paths \\
API scope & Narrow, stable API & Broad or extensible API \\
Ops & Footprint/deps matter & Ecosystem fit matters more \\
Validation & Oracle tests feasible & Oracle tests infeasible \\
\bottomrule
\end{tabular}
\setlength{\tabcolsep}{6pt}
\end{table}

In practice, three rules dominate: prefer stdlib-only replacements when
performance is within $2\times$ and deployment constraints matter; expect the
best wins where reference libraries carry architectural overhead; and keep the
third-party library when C-extension bottlenecks or broad API coverage dominate.
In all cases, reference-behavior validation remains essential before deployment.

\subsection{Limitations}

Several caveats apply. \zerodep targets utility and infrastructure libraries;
compute-intensive frameworks such as NumPy, PyTorch, or audio processing libraries
are outside its scope by design, as their performance is inseparable from
hardware-accelerated native code. Within that scope, \zerodep's API surface
sometimes omits advanced reference-library features (e.g., streaming iterators,
extension hooks), so some use cases will require manual extension. Subprocess
offloading recovers raw throughput but introduces platform variability and lacks
in-process guarantees such as thread safety and JIT-friendly data layouts. All
benchmarks ran on a single machine under controlled conditions; distributed,
multi-core, or GPU-accelerated environments may shift the performance balance
markedly.

Import-time memory was measured via \code{tracemalloc}, which instruments only
CPython's allocator and does not capture memory-mapped C-extension shared objects;
import-memory figures for C-extension-heavy references therefore underreport their
true footprint. Runtime memory under operational workloads was measured for six
representative modules but not systematically across the full collection.

\subsection{Threats to Validity}

\paragraph{Internal validity.}
Benchmarks ran on a single machine with pinned versions. \code{tracemalloc} underreports C-extension shared objects, and cold-start timing still includes residual process-startup noise despite repeated measurements. Runtime memory was measured for six representative modules rather than the full collection.

\paragraph{External validity.}
\zerodep targets utility and infrastructure libraries, not compute-intensive frameworks. Results may differ on other Python versions, operating systems, hardware architectures, or newer reference-library releases.

\paragraph{Construct validity.}
The $2\times$ parity threshold is a pragmatic engineering boundary rather than an empirically optimal one. Supply-chain metrics such as dependency count, maintainer count, and LOC are proxies for attack surface, not direct measures of exploitability.

\section{Conclusion}
\label{sec:conclusion}

We return to the question posed in our title: \emph{Stdlib or Third-Party?}
For the utility and infrastructure libraries in scope, the answer depends on
the workload, the deployment context, and the specification quality.
Roughly two-thirds of the modules achieve parity or better, and many outperform
their references when the reference design carries architectural overhead.
The main boundary is native acceleration: pure-Python stdlib replacements cannot
match C- or Rust-backed hot paths such as binary serialization, image processing,
and low-level cryptography.

For constrained LLM-assisted development, specification quality matters more than
model capability. Tighter contracts improve held-out compatibility, and
contract-linked review reduces review volume while preventing compatibility regressions.

Stdlib-only modules also offer clear deployment and trust-surface benefits: median
module size drops from 700\,KB to 31\,KB, cold-start latency improves by
$2.2\times$, and 88 transitive packages maintained by 128 distinct accounts are
eliminated. Taken together, these results position dependency elimination as an
evidence-based engineering tradeoff rather than an ad-hoc style preference.

\section*{Acknowledgments}

Large language models were used to assist with proofreading and language editing.
The author takes full responsibility for all content.

\bibliographystyle{unsrtnat}
\bibliography{references}

\end{document}